\documentclass[]{memtensor}
\usepackage{enumitem}
\usepackage[utf8]{inputenc}
\usepackage{url}
\usepackage{graphicx}
\usepackage{booktabs}
\usepackage{amsfonts}
\usepackage{nicefrac}
\usepackage{makecell}
\usepackage{microtype}
\usepackage{amsmath}
\usepackage{etoolbox}
\usepackage{lipsum}
\usepackage{minitoc}
\usepackage{tablefootnote}
\usepackage{threeparttable}
\usepackage{wrapfig}
\usepackage{appendix}
\usepackage{multirow}
\usepackage{ulem}
\useunder{\uline}{\ul}{}
\usepackage{colortbl}
\usepackage{adjustbox}
\usepackage{array}
\usepackage{tabularx}
\usepackage{siunitx}
\usepackage{placeins}
\usepackage{float}
\usepackage{xcolor}
\usepackage{mdframed}
\usepackage{listings}

\graphicspath{{img/}}
\IfFileExists{fontawesome5.sty}{
    \usepackage{fontawesome5}
}{
    
    \newcommand{\faGithub}{[GH]}

}

\newcommand{\iconbox}[1]{\makebox[1.5em][c]{\large #1}}

\newcommand{\hficon}{\raisebox{-0.15em}{\IfFileExists{logo/huggingface.png}{\includegraphics[height=1.1em]{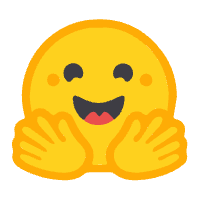}}{[HF]}}}

\newcommand{\figplaceholder}[1]{%
  \fbox{%
    \parbox[c][0.23\textheight][c]{0.9\linewidth}{\centering #1}%
  }%
}

\newcommand{\smartincludegraphics}[2]{%
  \IfFileExists{img/#1}{\includegraphics[width=#2]{#1}}{\figplaceholder{Missing figure: img/#1}}%
}

\lstdefinestyle{promptlisting}{
  basicstyle=\ttfamily\scriptsize,
  breaklines=true,
  breakatwhitespace=false,
  columns=fullflexible,
  keepspaces=true,
  showstringspaces=false,
  upquote=true,
  frame=none
}

\mdfdefinestyle{promptboxstyle}{
  linecolor=gray!60,
  linewidth=0.8pt,
  backgroundcolor=gray!8,
  roundcorner=2pt,
  innertopmargin=8pt,
  innerbottommargin=6pt,
  innerleftmargin=8pt,
  innerrightmargin=8pt,
  skipabove=8pt,
  skipbelow=8pt,
  frametitleaboveskip=0.5em,
  frametitlebelowskip=0.5em,
  frametitlerule=true,
  frametitlerulewidth=0.8pt,
  frametitlebackgroundcolor=gray!20,
  frametitlefont=\bfseries\small
}

\newenvironment{promptbox}[1]{
  \begin{mdframed}[style=promptboxstyle,frametitle={#1}]
}{
  \end{mdframed}
}

\usepackage{tikz}
\tcbuselibrary{skins, breakable}

\definecolor{qborder}{HTML}{378ADD}
\definecolor{qbg}{HTML}{E6F1FB}
\definecolor{thinkborder}{HTML}{AFA9EC}
\definecolor{thinkbg}{HTML}{F5F4FE}
\definecolor{retainedbg}{HTML}{EAF3DE}
\definecolor{retainedtext}{HTML}{27500A}
\definecolor{retainedborder}{HTML}{97C459}
\definecolor{discardedbg}{HTML}{F1EFE8}
\definecolor{discardedtext}{HTML}{888780}
\definecolor{answerbg}{HTML}{EAF3DE}
\definecolor{answerborder}{HTML}{639922}
\definecolor{serialbg}{HTML}{E6F1FB}
\definecolor{serialtext}{HTML}{0C447C}
\definecolor{parallelbg}{HTML}{E1F5EE}
\definecolor{paralleltext}{HTML}{085041}
\definecolor{reflectbg}{HTML}{FAECE7}
\definecolor{reflecttext}{HTML}{712B13}

\newtcolorbox{casebox}[1]{
  enhanced, breakable,
  colback=white, colframe=gray!30,
  fonttitle=\bfseries\small,
  title={#1},
  boxrule=0.4pt, arc=4pt,
  top=6pt, bottom=6pt, left=8pt, right=8pt,
  toptitle=4pt, bottomtitle=4pt,
}

\newtcolorbox{questionbox}{
  enhanced,
  colback=qbg, colframe=qborder,
  boxrule=0pt, leftrule=3pt, arc=0pt,
  top=4pt, bottom=4pt, left=8pt, right=8pt,
}

\newtcolorbox{thinkbox}{
  enhanced,
  colback=thinkbg, colframe=thinkborder,
  boxrule=0pt, leftrule=2pt, arc=0pt,
  top=3pt, bottom=3pt, left=6pt, right=6pt,
  fontupper=\small\itshape\color{gray!70!black},
}

\newtcolorbox{answerbox}{
  enhanced,
  colback=answerbg, colframe=answerborder,
  boxrule=0pt, leftrule=3pt, arc=0pt,
  top=4pt, bottom=4pt, left=8pt, right=8pt,
}

\newcommand{\tagserial}{%
  \tikz[baseline=(tag.base)]{\node[fill=serialbg, text=serialtext,
    rounded corners=2pt, inner sep=2pt, font=\scriptsize\bfseries] (tag) {serial search};}%
}
\newcommand{\tagparallel}{%
  \tikz[baseline=(tag.base)]{\node[fill=parallelbg, text=paralleltext,
    rounded corners=2pt, inner sep=2pt, font=\scriptsize\bfseries] (tag) {parallel search};}%
}
\newcommand{\tagreflect}{%
  \tikz[baseline=(tag.base)]{\node[fill=reflectbg, text=reflecttext,
    rounded corners=2pt, inner sep=2pt, font=\scriptsize\bfseries] (tag) {reflect \& denoise};}%
}

\newcommand{\evretained}[2]{%
  \par\noindent\colorbox{retainedbg}{%
    \parbox{\dimexpr\linewidth-2\fboxsep}{%
      \small\color{retainedtext}\textbf{[#1]}~#2%
    }%
  }\vspace{1pt}%
}
\newcommand{\evdiscarded}[2]{%
  \par\noindent\colorbox{discardedbg}{%
    \parbox{\dimexpr\linewidth-2\fboxsep}{%
      \small\color{discardedtext}\sout{[#1]~#2}%
    }%
  }\vspace{1pt}%
}

\newcommand{\queryitem}[1]{%
  \par\noindent{\small\ttfamily\color{gray!70!black}#1}\vspace{1pt}%
}

\title{%
  \titlefont%
  \parbox{\textwidth}{%
    \centering
    {\bfseries MemRetriever: Learning to Search, Reflect, and Retrieve from Long-Term Memory} %
  }%
}

\author[1,*]{Ruiyang Jiang}
\author[1,*]{Chunyu Li}
\author[1]{Zhiyu Li}

\affiliation[1]{MemTensor (Shanghai) Technology} 
\affiliation[*]{Equal Contribution}

\abstract{
Long-term memory is essential for personalized agents, but effective memory use depends not only on storing information, but also on retrieving the right evidence at the right time. Existing memory retrieval systems commonly rely on static top-$k$ retrieval, which issues a single query, returns a fixed number of memories, and directly passes them to a downstream model. This paradigm is often insufficient for complex questions that require multi-hop reasoning, temporal comparison, knowledge updates, or evidence scattered across multiple sessions; it may miss key memories, introduce noisy snippets, and waste context budget. In this paper, we introduce MemRetriever, an agentic memory retrieval model that formulates memory access as a multi-step search process. Instead of passively accepting one-shot retrieval results, MemRetriever reasons over the current evidence state, actively chooses among parallel search, serial search, and reflection-based denoising, and stops when the retained evidence is sufficient for downstream answering. We construct ReAct-style search-memory trajectories for supervised warm start and further optimize the model with Group Relative Policy Optimization (GRPO), using reward signals that encourage evidence coverage, noise reduction, and answer sufficiency. Experiments on LOCOMO, LongMemEval, HotpotQA, MuSiQue, and 2WikiMultiHopQA show that MemRetriever improves both long-term memory retrieval and complex question answering over static retrieval baselines and supervised-only variants. On LongMemEval, MemRetriever-4B-RL surpasses DeepSeek-v4-Flash under the same retrieval pipeline on the main retrieval metrics. On multi-hop QA benchmarks, MemRetriever also achieves strong results, including the best reported EM/F1/Judge scores among the compared methods on MuSiQue. Beyond memory stores, MemRetriever models storage-agnostic retrieval decision logic and can also be applied to external knowledge bases or vector databases, making it a more general retrieval model for retrieval-augmented agent systems. The results suggest that long-term agent memory and broader knowledge-intensive applications benefit from an intermediate decision layer that can plan when to search, what to retrieve, how to filter evidence, and when to stop.

\vspace{1.5em}
\noindent
\begingroup

\begin{tabular}{@{}l@{}}
\iconbox{\faEnvelope} \textbf{Corresponding Author:} Zhiyu Li (lizy@memtensor.cn)\\
\iconbox{\faGithub} \textbf{Github:} \href{https://github.com/MemTensor/MemOS}{github.com/MemTensor/MemOS} \\
\iconbox{\hficon} \textbf{MemRetriever-4B:} \href{}{} \\
\end{tabular}
\endgroup
}

\begin{document}
\maketitle

\newpage

\section{Introduction}
LLM-driven agents continuously accumulate dialogue histories through ongoing interactions with users, and these histories constitute the agent’s long-term memory. When a user asks a question that requires recalling past information—for example, “What was the name of the restaurant I mentioned last time?” or “How many volunteer activities have I participated in altogether?”—the agent must retrieve relevant evidence from historical conversations that may span dozens or even hundreds of turns in order to answer accurately.

However, most existing memory retrieval methods adopt a static top-k paradigm: given a user query, they perform a one-shot retrieval of the k most relevant memory snippets based on embedding similarity, and directly concatenate them into the context for the model to answer. This paradigm has three fundamental limitations. First, a single query often fails to cover all the evidence required for multi-hop questions. When an answer depends on multiple facts scattered across different conversations, a single query vector usually retrieves only part of the evidence, resulting in incomplete recall. Second, a fixed value of k cannot adapt to the varying evidence needs of different questions: a simple question may require only one memory, whereas a complex question may require more than ten. Static truncation therefore either introduces noise or omits key information. Third, retrieved results lack a quality-filtering mechanism, causing noisy and redundant documents to be mixed with genuinely useful evidence. This increases the reasoning burden on the downstream model and reduces answer accuracy.

In recent years, agentic search methods have shown significant advantages in open-domain question answering. Work represented by ReAct demonstrates that explicitly decoupling reasoning and action into a think–action–observation loop enables a model to dynamically adjust subsequent search strategies based on intermediate retrieval results, thereby effectively handling multi-hop reasoning and complex information needs. However, existing agentic search work mainly targets open-domain question answering or external document retrieval, where the search space consists of encyclopedic or web corpora. This differs substantially from the agent memory setting: information in a memory store is more fine-grained, spans a longer time range, and the same entity may be mentioned in different ways across different conversations.

This work proposes a method for introducing agentic search into long-term memory retrieval for agents. We formulate memory retrieval as a multi-step search process: before answering a question, a dedicated search agent actively searches, filters, and completes supporting evidence from the memory store through multiple rounds of think–act–observe loops, ultimately constructing a structured evidence pool for the downstream answering model.

Specifically, in each loop, the search agent first reasons, or “thinks,” about the current state of the evidence pool and the remaining information gaps. It then selects one of three actions to execute: parallel search, serial search, or reflection and denoising. Parallel search issues multiple queries simultaneously to explore different retrieval directions, making it suitable for scenarios where the evidence space is ambiguous or the information need is multifaceted. Serial search issues a single precise query to fill a specific missing piece of information, making it suitable when a concrete information gap has already been identified. Reflection and denoising scores, filters, and deduplicates the candidate evidence in the current evidence pool, removes noisy documents, and determines whether the evidence is sufficient. When reflection determines that the evidence is sufficient, the search loop terminates; otherwise, retrieval continues until the maximum number of rounds is reached.

To train the search agent’s decision-making ability through reinforcement learning, we design a set of reward functions aligned with the multi-step search process. The search reward is based on incremental coverage gain, which measures the contribution of each retrieval step to the coverage of gold evidence and provides a dense training signal. The reflection reward adopts a coverage-preserving criterion, strongly penalizing the mistaken removal of useful evidence while mildly rewarding the removal of redundancy. The termination reward uses a downstream reader model to evaluate the answer sufficiency of the final evidence pool, linking the quality of the entire search trajectory to the final answering performance.

It is worth emphasizing that although this paper uses memory retrieval as the core validation scenario, what we model is essentially storage-agnostic retrieval decision logic: when to retrieve, how to retrieve, and when to stop. As an independent intermediate decision layer, the search agent can be deployed either between the model and the memory store or between the model and an external vector knowledge base such as Milvus, dynamically planning retrieval behavior according to the current dialogue state. Therefore, the proposed method is not limited to memory scenarios and has the potential to transfer to a broader range of retrieval-augmented generation tasks.

The main contributions of this paper are as follows:
\begin{itemize}
    \item We formulate long-term memory retrieval for agents as a multi-step agentic search process and propose a search-agent framework composed of three operations: parallel search, serial search, and reflection-based denoising.
    \item We design a hierarchical reward function aligned with the multi-step search process: the search stage uses incremental coverage metrics to encourage exploration, the reflection stage uses recall-preserving coverage constraints, and the termination stage evaluates evidence sufficiency through a downstream reader model. 
    \item Experiments on QA datasets such as HotpotQA and memory datasets such as LOCOMO show that the proposed method achieves significant improvements across multiple metrics over static top-$k$ retrieval baselines.
\end{itemize}

\section{Related Work}

\subsection{Agentic Search}

Retrieval-augmented generation (RAG) mitigates the knowledge-boundary and hallucination problems of large language models by introducing external knowledge bases~\cite{lewis2020retrieval}. Traditional RAG typically adopts embedding-based retrieval, where documents relevant to the input question are concatenated into the context before answer generation. However, this static retrieval pipeline struggles with complex questions that require multi-step evidence acquisition and query rewriting. As a result, subsequent work began to treat search as a tool callable by the model, enabling the model to alternate between thinking and retrieval during the reasoning process.

Search-R1 is an early work that applies reinforcement learning to agentic search~\cite{jin2025searchr1}. It models search as part of an RL environment, allowing the model to alternately generate reasoning content and search queries during rollout, and to continue reasoning based on the returned evidence. This method stabilizes training by masking tokens returned from retrieval and mainly relies on final-answer rewards to optimize the model’s retrieval-augmented reasoning ability. ReSearch further treats search operations as components of the reasoning chain~\cite{chen2503research}. Using GRPO without supervised search trajectories, it trains the model to learn when to search, how to search, and how to use retrieved results to continue reasoning, making it the main agentic search method compared in this paper. More recent work studies the retrieval side of agentic search: Agentic-R trains retrievers with both local passage relevance and global answer correctness for multi-turn search agents~\cite{liu2026agenticr}, while SIGHT uses self-evidence support and information-gain-driven branching to reduce redundancy and low signal-to-noise observations in multi-turn search~\cite{zhong2026sight}.

Although these methods demonstrate the effectiveness of RL for training search agents, most of them rely on final-answer rewards, which leads to sparse rewards and unclear credit assignment for intermediate search steps. MR-Search improves the search process from the perspective of cross-episode reflection~\cite{xiao2026meta}, enabling the model to generate self-reflections based on previous attempts and gradually revise its strategy in subsequent searches. Together, these methods advance agentic search from static retrieval toward an interactive, process-aware, and reflection-capable search paradigm.

\subsection{Memory Retrieval in Agent Memory}

Memory is an important mechanism that enables LLM agents to maintain long-term interaction and personalization capabilities~\cite{ai_memory_survey}. Early memory-system work such as MemGPT frames long-term interaction as virtual context management across memory tiers~\cite{packer2023memgpt}, while MemoryBank introduces a long-term memory mechanism that stores, updates, and recalls user interaction history for personalized LLM companions~\cite{zhong2024memorybank}. Existing agent memory methods typically store historical interactions as external memory and retrieve relevant snippets when answering the current question. However, as the memory scale grows, simple top-$k$ similarity retrieval can easily lead to efficiency degradation and noise interference.

SwiftMem focuses on the efficiency problem in memory retrieval~\cite{tian2026swiftmem}, pointing out that many memory frameworks perform exhaustive retrieval over the entire memory storage, resulting in high latency. To address this issue, it proposes a query-aware temporal index and a semantic DAG-tag index, enabling the system to retrieve only relevant memory subsets according to query characteristics. MemSifter, by contrast, delegates memory retrieval to a lightweight proxy model and trains this model through outcome-driven RL, so that the selected memories can genuinely improve the downstream task performance of the working LLM~\cite{tan2026memsifter}.

These methods improve the efficiency or task relevance of long-term memory retrieval, but they still typically model memory access as an indexing, ranking, or single-stage filtering problem. They lack an explicit multi-step search process for determining whether the current evidence is sufficient, whether information is missing, and which memories should be retained or removed.

\subsection{Reasoning-Augmented and Tool-Use Agents}

ReAct, Toolformer, and a large body of subsequent work on tool-augmented agents~\cite{react,toolformer} show that explicitly decoupling “reasoning” from “action” can significantly improve the robustness and interpretability of models on complex tasks. In particular, in scenarios involving external retrieval, state tracking, multi-step decision-making, or deferred judgment, an explicit think–action–observation loop often outperforms one-shot end-to-end generation. A complementary line of work studies reflection or self-reflection as an agentic feedback mechanism, where the model critiques previous observations or decisions, identifies failure modes, and uses the resulting feedback to refine subsequent actions~\cite{reflexion}. This paradigm has now demonstrated strong effectiveness across multiple directions, including task planning, code generation, knowledge retrieval, and interactive agents.

This paper draws inspiration from this paradigm, but its focus is not general-purpose task solving. Instead, we focus on action decision-making within the retrieval process. We instantiate tool use as three operations: serial search, parallel search, and reflection and denoising. The resulting search agent serves as an independent intermediate decision layer, enabling it to interface with different types of backend storage systems and retrieval objectives.

\subsection{Our Contributions}

Compared with the above work, this paper focuses on introducing agentic search into the agent memory setting. Existing agentic search methods mainly target open-domain question answering or external document retrieval, while existing memory retrieval methods mostly focus on index structures, ranking strategies, or context compression. MemSearcher is an early attempt to connect these two research lines~\cite{yuan2025memsearcher}. It maintains a compact memory to prevent a ReAct-style search agent from continuously concatenating the full interaction history, thereby effectively reducing context length and noise. However, MemSearcher mainly addresses the problem of accumulated interaction-history expansion during search, whereas this paper focuses on a more upstream problem: before generating an answer, how can an agent actively search, filter, and complete supporting evidence from long-term memory?

To this end, this paper formulates memory retrieval as a multi-step agentic search process. The proposed search agent does not directly answer the user’s question; instead, it constructs a structured memory evidence pool for a downstream answering model. Specifically, when the evidence space is ambiguous or the retrieval direction is unclear, the model performs parallel search to explore multiple potential directions simultaneously. When a specific information gap has been identified, the model switches to serial search to complete the evidence in a targeted manner. When the retrieved results contain noise or redundancy, the model scores, filters, and updates candidate evidence through reflection and denoising. Through this mechanism, this paper advances retrieval in agent memory from static top-$k$ selection to a dynamic process of evidence discovery and evidence purification.

Furthermore, although this paper uses memory retrieval as the core validation scenario, what we model is essentially storage-agnostic retrieval decision logic: when to retrieve, how to retrieve, and when to stop. As an independent intermediate decision layer, the search agent can be deployed either between the model and the memory store or between the model and an external vector knowledge base such as Milvus, dynamically planning retrieval behavior according to the current dialogue state. Therefore, the proposed method is not limited to memory scenarios and has the potential to transfer to a broader range of retrieval-augmented generation tasks.

\section{Method}
\label{sec:problem}

Following the ReAct interaction paradigm~\cite{react}, we formulate memory retrieval as a multi-step agentic search problem rather than a one-shot top-$k$ selection problem. Given a user question $q$ and a long-term memory store $\mathcal{M}$, the goal of the search agent is not to directly answer $q$, but to construct a compact and sufficient evidence pool $\mathcal{E}$ for a downstream reader model.

\begin{figure}[t]
    \centering
    \smartincludegraphics{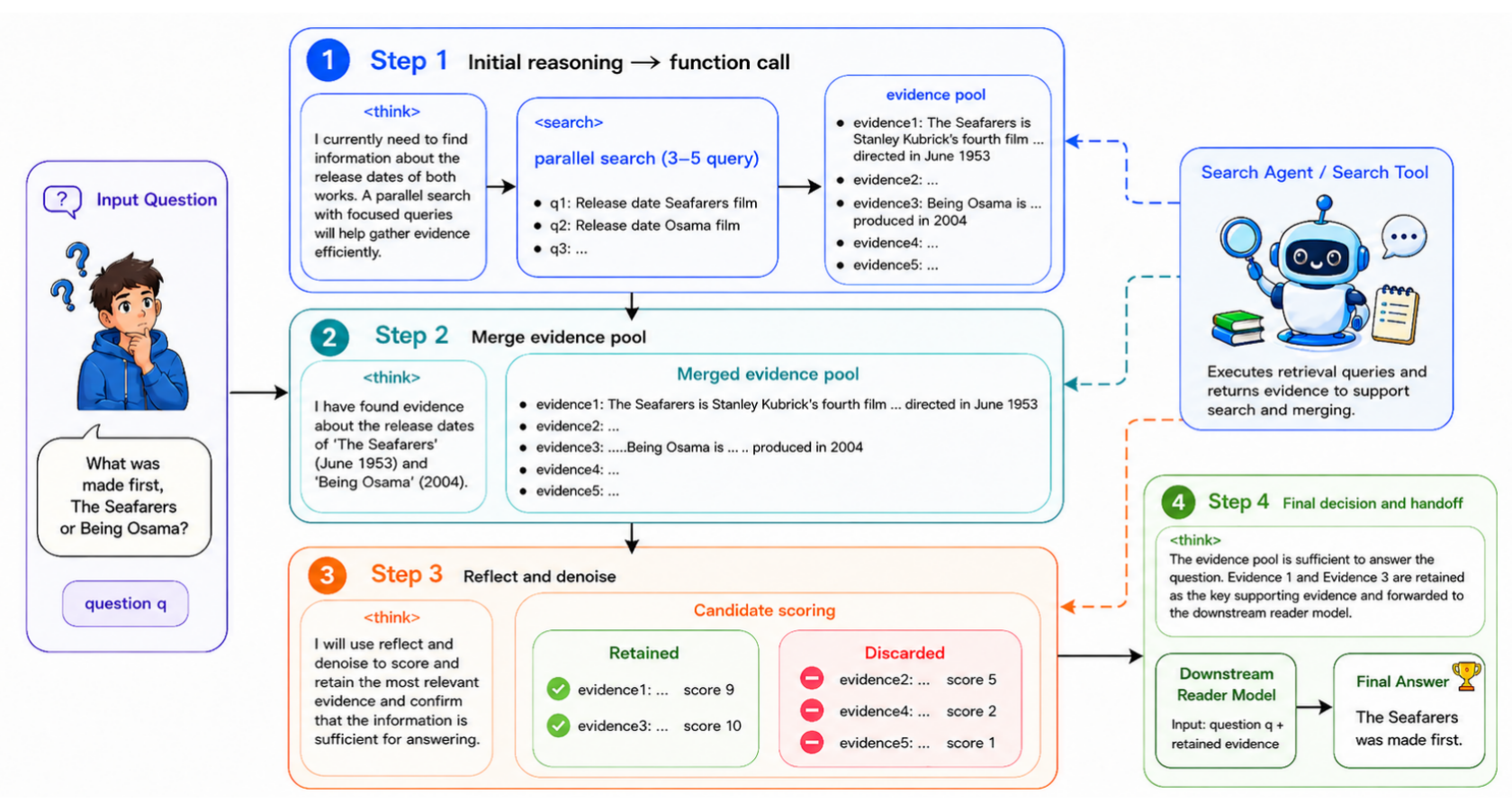}{0.95\linewidth}
    \caption{\textbf{Overall workflow of MemRetriever.} Given a user question and a long-term memory store, the search agent iteratively reasons over the current evidence pool, invokes parallel search, serial search, or reflection-based denoising, and stops when the collected evidence is sufficient for downstream answer generation.}
    \label{fig:workflow}
\end{figure}

At step $t$, the agent observes the question, the current evidence pool, and the previous search history. We define the decision state as
\begin{equation}
    s_t = (q, \mathcal{M}, \mathcal{E}_{t-1}, \mathcal{H}_{t-1}),
\end{equation}
where $\mathcal{E}_{t-1}$ denotes the evidence pool collected before step $t$, and $\mathcal{H}_{t-1}$ denotes the previous think--action--observation history.

Given $s_t$, MemRetriever generates a ReAct-style search trajectory
\begin{equation}
    \tau = \{(z_t, a_t, o_t)\}_{t=1}^{T}, \qquad a_t \in \mathcal{A},
\end{equation}
where $z_t$ denotes the reasoning trace that analyzes the current evidence state and remaining information gaps, $a_t$ is the selected retrieval action, and $o_t$ is the observation returned by the retrieval backend or the reflection module. Consistent with the retrieval operations introduced in Section~2.4, the action space is defined as
\begin{equation}
    \mathcal{A}
    =
    \{\texttt{parallel\_search},\ \texttt{serial\_search},\ \texttt{reflection\_and\_denoise}\}.
\end{equation}
Here, \texttt{parallel\_search} issues multiple queries to explore different retrieval directions when the evidence space is ambiguous; \texttt{serial\_search} issues a targeted query to fill a specific missing fact; and \texttt{reflection\_and\_denoise} scores, filters, and deduplicates the current candidates. In addition, \texttt{reflection\_and\_denoise} returns a JSON field \texttt{done} indicating whether the retained evidence is sufficient for downstream answering; when this field is set to true, \texttt{stop} terminates the search process and returns the final evidence pool.

The evidence pool is updated by a transition operator
\begin{equation}
    \mathcal{E}_t = \mathcal{T}(\mathcal{E}_{t-1}, a_t, o_t),
\end{equation}
where search actions expand $\mathcal{E}$ with newly retrieved memory snippets, while reflection actions remove noisy or redundant candidates, preserve useful evidence, and may set the JSON field \texttt{done} to true. When \texttt{done} is true, the environment terminates the search loop and returns the final evidence pool $\mathcal{E}_T$ to the downstream reader.

We optimize the search policy $\pi_\theta(\cdot\mid s_t)$ to maximize the expected utility of the complete retrieval trajectory:
\begin{equation}
    J(\theta)
    =
    \mathbb{E}_{\tau \sim \pi_\theta}
    \left[
    \sum_{t=1}^{T} \gamma^{t-1} R(s_t, a_t, \mathcal{E}_t)
    \right],
\end{equation}
where $R(s_t, a_t, \mathcal{E}_t)$ combines the search, reflection, stop, and format rewards described in Section~3.2. This objective encourages the agent to learn when to retrieve, how to retrieve, how to denoise, and when to stop.

\subsection{Supervised Warm Start}

We first initialize the policy model using supervised trajectories. Let the supervised fine-tuning dataset be denoted as $\mathcal{D}_{\mathrm{sft}}$, where each sample consists of an input $x$ and a target token sequence
$y = (y_1, \ldots, y_T)$. The target sequence contains the complete output trajectory expected by the task, including the reasoning process enclosed by \texttt{<think>} and \texttt{</think>}, tool calls in the form of \texttt{<function\_call>} and \texttt{</function\_call>}, and the corresponding structured arguments. In our setting, the SFT stage requires the model not only to learn search trajectories, but also to follow the predefined output protocol, such as producing well-formed \texttt{<think>} traces, valid JSON tool calls, and the basic semantics of the three types of search tools.

We optimize the model with the standard next-token maximum likelihood objective:
\begin{equation}
\mathcal{L}_{\mathrm{SFT}}(\theta)
=
-\sum_{t=1}^{T}
\log \pi_{\theta}\left(y_t \mid x, y_{<t}\right),
\end{equation}
where $\pi_{\theta}$ denotes the policy model to be optimized, and $y_{<t}$ denotes the prefix tokens before position $t$. This objective encourages the model to imitate the supervised trajectory token by token, thereby learning the basic output format and task behavior before entering the reinforcement learning stage.

The purpose of this warm-start stage is to provide a stable initial policy for subsequent reinforcement learning. Since the RL stage relies on the model to generate complete trajectories that are both parseable and evaluable, a model that has not yet mastered the basic protocol may frequently produce formatting errors, invalid JSON arguments, incomplete tool calls, or incorrect use of search actions. These issues can increase noise in reward evaluation and reduce the proportion of valid training samples. By using SFT to pretrain the output protocol and the basic semantics of the three search tools, the model can generate more stable candidate trajectories during reinforcement learning, allowing subsequent GRPO optimization to focus more on improving decision quality and policy preferences rather than relearning basic formatting.

\subsection{Reward Design for Memory Retrieval}

The ultimate goal of MemRetriever is to collect sufficient and effective evidence that can support the downstream model in generating the correct answer. However, relying solely on final-answer correctness as the reward leads to a severe reward sparsity problem. Since MemRetriever must go through multiple rounds of search, evidence filtering, and information integration before receiving final feedback, credit assignment during training becomes difficult. To address this issue, we decompose the overall retrieval process into three stages: \textit{search}, \textit{reflection}, and \textit{stop}, and design stage-specific reward signals accordingly. In addition, to ensure stable tool use and reasoning trajectories, we introduce an extra format penalty term.

Let a search trajectory be denoted as
\begin{equation}
\tau = \{(s_t, a_t, o_t)\}_{t=1}^{T},
\end{equation}
where $s_t$ denotes the current state, $a_t$ denotes the action selected by the search agent, and $o_t$ denotes the observation returned by the retrieval environment or tool execution. The total reward is defined as
\begin{equation}
R(\tau)
=
\sum_{t=1}^{T}
\left(
R_{\mathrm{search}}^{(t)}
+
R_{\mathrm{reflect}}^{(t)}
\right)
+
R_{\mathrm{stop}}
-
R_{\mathrm{format}}.
\end{equation}

\paragraph{Search Reward.}
The primary objective of the search stage is to continuously expand the evidence pool and improve the coverage of question-relevant facts. Therefore, we use coverage gain as the core search reward, rather than simply rewarding the number of retrieved documents. Specifically, we compute the change in coverage of the gold evidence set before and after each search step, and use this incremental gain as the reward signal:
\begin{equation}
R_{\mathrm{search}}^{(t)}
=
\mathrm{Cov}(E_t, G)
-
\mathrm{Cov}(E_{t-1}, G)
-
\lambda_q C_q(a_t),
\end{equation}
where $E_t$ denotes the evidence pool after step $t$, $G$ denotes the gold evidence set, $\mathrm{Cov}(\cdot)$ measures the coverage of gold evidence, $C_q(a_t)$ is the query cost of action $a_t$, and $\lambda_q$ controls the strength of the query-cost penalty.

This coverage-gain design has two advantages. First, it encourages the agent to retrieve documents that provide new information, rather than repeatedly retrieving content already present in the evidence pool, thereby improving evidence diversity. Second, the incremental reward provides continuous feedback throughout the search process, effectively alleviating reward sparsity in reinforcement learning. In addition, we apply a small query-cost penalty to each search request. Without such a cost constraint, the model may perform many unnecessary searches to accumulate coverage. The query penalty forces the model to trade off information gain against search cost, leading to more efficient search strategies.

\paragraph{Reflection Reward.}
As the number of search rounds increases, the evidence pool inevitably accumulates redundant information and low-value content. We therefore introduce a reflection module to filter, organize, and deduplicate the collected evidence, thereby improving the information density of the evidence pool.

In the early stage of reward design, we adopted a similarity-based matching mechanism using token-level F1 and semantic similarity to evaluate reflection quality. This design assigns graded rewards according to the similarity between the retained evidence after reflection and the gold evidence. However, experiments showed that this design has clear limitations. Since F1 and semantic similarity mainly capture surface-level textual closeness rather than factual coverage, many pieces of evidence that are highly similar to the final evidence but do not contain key facts can still receive high rewards. As a result, the model tends to retain many ``seemingly relevant'' memory snippets that cannot effectively support the final answer.

Further analysis shows that the evidence pool gradually accumulates a large amount of high-similarity but low-value information. Although local rewards continue to increase, final question-answering performance does not improve accordingly. This indicates that similarity-based rewards cannot effectively distinguish evidence that truly covers the final facts from evidence that is merely semantically similar, causing a mismatch between the reward objective and the final task objective.

Based on this observation, we adopt a coverage-based validation mechanism grounded in gold evidence coverage. Only evidence that covers final facts is considered valid and participates in reward computation. Compared with similarity-derived scores, this design directly ties the reward to key facts and significantly reduces the accumulation of noisy memories and pseudo-relevant information.

However, after adopting coverage-based rewards, we observe another problem: the reflection module may mistakenly remove important evidence that already covers gold facts during deduplication. To address this issue, we introduce a gold-evidence loss penalty. Let $E_t^{-}$ and $E_t^{+}$ denote the evidence pool before and after reflection at step $t$, respectively. We first define the set of gold facts covered by an evidence pool as
\begin{equation}
\mathcal{C}(E, G)
=
\{g \in G \mid \exists e \in E,\; \mathrm{cover}(e,g)=1\},
\end{equation}
where $\mathrm{cover}(e,g)$ is a binary indicator of whether evidence $e$ supports gold fact $g$. The number of lost gold facts after reflection is then computed as
\begin{equation}
n_{\mathrm{drop}}^{(t)}
=
\left|
\mathcal{C}(E_t^{-},G)
\setminus
\mathcal{C}(E_t^{+},G)
\right|.
\end{equation}
To avoid rewarding aggressive deletion, we give a redundancy bonus only when reflection removes evidence without reducing gold coverage. We define the number of safely removed redundant evidence items as
\begin{equation}
n_{\mathrm{redundant}}^{(t)}
=
\mathbb{I}\left[n_{\mathrm{drop}}^{(t)}=0\right]
\cdot
\max\left(0, |E_t^{-}|-|E_t^{+}|\right).
\end{equation}
The reflection reward is therefore formulated as
\begin{equation}
R_{\mathrm{reflect}}^{(t)}
=
\begin{cases}
-\lambda_{\mathrm{drop}} n_{\mathrm{drop}}^{(t)},
& n_{\mathrm{drop}}^{(t)} > 0, \\
\min\left(r_{\max},\lambda_{\mathrm{bonus}} n_{\mathrm{redundant}}^{(t)}\right),
& n_{\mathrm{drop}}^{(t)} = 0.
\end{cases}
\end{equation}
Equivalently, this can be written in a unified form as
\begin{equation}
R_{\mathrm{reflect}}^{(t)}
=
\max\left(
r_{\min},
-\lambda_{\mathrm{drop}} n_{\mathrm{drop}}^{(t)}
+
\mathbb{I}\left[n_{\mathrm{drop}}^{(t)}=0\right]
\cdot
\min\left(r_{\max},\lambda_{\mathrm{bonus}} n_{\mathrm{redundant}}^{(t)}\right)
\right).
\end{equation}
We set $\lambda_{\mathrm{drop}} = 0.6$, $\lambda_{\mathrm{bonus}} = 0.1$, $r_{\max} = 0.3$, and $r_{\min} = -1.5$.
Here, $\lambda_{\mathrm{drop}}$ controls the penalty for dropping gold evidence, $\lambda_{\mathrm{bonus}}$ controls the reward for safely removing redundant evidence, and $r_{\max}$ and $r_{\min}$ bound the reflection reward.

\paragraph{Stop Reward.}
Both search and reflection ultimately serve the final question-answering task. Therefore, relying only on intermediate rewards cannot guarantee that the collected evidence truly supports problem solving. To this end, we introduce a stop reward as the global optimization objective. When the agent decides to stop searching, a downstream reader model generates an answer based on the final evidence pool, and the generated answer is compared with the reference answer. The stop reward is directly determined by answer quality:
\begin{equation}
R_{\mathrm{stop}}
=
\mathrm{Score}(\hat{y}, y^{*}),
\end{equation}
where $\hat{y}$ denotes the answer generated by the downstream reader model, $y^{*}$ denotes the reference answer, and $\mathrm{Score}(\cdot)$ measures answer quality.

This design encourages the agent to focus not only on how much information is retrieved, but also on whether the retrieved information can support correct reasoning and answer generation. In other words, the search and reflection rewards provide local optimization signals, while the stop reward provides the final task-level objective. Moreover, the stop reward helps the model learn a reasonable stopping policy. When the current evidence is already sufficient to support a correct answer, the marginal benefit of further search decreases, while additional search costs continue to accumulate. As a result, the model can learn to stop when the evidence is sufficient, rather than unnecessarily extending the search trajectory. In our experiments, we use GPT-4.1-mini as the downstream reader model. 

\paragraph{Format Penalty.}
Since the agent frequently performs tool calls and structured reasoning, formatting errors in the trajectory may cause tool execution failures or abnormal training signals. To improve training stability, we introduce an additional format penalty for invalid tool calls, mismatched tags, malformed structures, and outputs that violate the predefined protocol:
\begin{equation}
R_{\mathrm{format}}
=
\min
\left(
N_{\mathrm{err}} \cdot p_{\mathrm{format}},
p_{\max}
\right),
\end{equation}
where $N_{\mathrm{err}}$ denotes the total number of format errors in the trajectory, $p_{\mathrm{format}}$ is the penalty for each format error, and $p_{\max}$ is the maximum penalty. In our implementation, we set $p_{\mathrm{format}} = 0.05$ and $p_{\max} = 0.3$.

Although the format penalty accounts for only a small portion of the total reward, it substantially reduces invalid trajectories, improves training stability, and increases the success rate of tool invocation.

\subsection{GRPO Objective}

Starting from the SFT-initialized policy, we further optimize the search agent with GRPO~\cite{deepseekmath}, a PPO-style reinforcement learning method that avoids training a separate critic~\cite{ppo}. For each retrieval state $s$, we sample a group of $G$ candidate search trajectories $\{y_i\}_{i=1}^{G}$ from the old policy $\pi_{\theta_{\mathrm{old}}}(\cdot \mid s)$. Following the reward design above, each candidate trajectory is first scored by the complete memory-retrieval reward:
\begin{equation}
R_i
=
R(y_i)
=
\sum_{t=1}^{T_i}
\left(
R_{\mathrm{search},i}^{(t)}
+
R_{\mathrm{reflect},i}^{(t)}
\right)
+
R_{\mathrm{stop},i}
-
R_{\mathrm{format},i}.
\end{equation}
The group-relative advantage is then computed from these trajectory-level rewards:
\begin{equation}
    \hat{A}_i
    =
    \frac{R_i - \mathrm{mean}\left(\{R_j\}_{j=1}^{G}\right)}
    {\mathrm{std}\left(\{R_j\}_{j=1}^{G}\right) + \epsilon}.
\end{equation}
where $\epsilon > 0$ is a small numerical-stability constant used to avoid division by zero. For each token position $\ell$ in candidate trajectory $y_i$, we define the importance ratio as
\begin{equation}
    \rho_{i,\ell}(\theta) = \frac{\pi_\theta(y_{i,\ell}\mid s, y_{i,<\ell})}{\pi_{\theta_{\mathrm{old}}}(y_{i,\ell}\mid s, y_{i,<\ell})}.
\end{equation}
The clipped GRPO objective is given by
\begin{equation}
\mathcal{J}_{\mathrm{GRPO}}(\theta) =
\mathbb{E}_{s,\{y_i\}}\!\Bigg[\frac{1}{G}\sum_{i=1}^{G}\frac{1}{|y_i|}\sum_{\ell=1}^{|y_i|}
\min\!\Big(\rho_{i,\ell}(\theta)\hat{A}_i,\;
\operatorname{clip}(\rho_{i,\ell}(\theta), 1{-}\epsilon_c, 1{+}\epsilon_c)\hat{A}_i\Big)\!\Bigg]
- \beta\,D_{\mathrm{KL}}\!\big(\pi_\theta \| \pi_{\mathrm{ref}}\big),
\end{equation}
where $\epsilon_c$ denotes the clipping range, $\pi_{\mathrm{ref}}$ denotes the reference policy used for KL regularization, and $\beta$ controls the strength of the KL penalty.

Training follows a two-stage pipeline. We first learn an SFT-initialized policy by solving
\begin{equation}
    \theta_{\mathrm{SFT}} = \arg\min_{\theta} \mathcal{L}_{\mathrm{SFT}}(\theta).
\end{equation}
Starting from $\theta_{\mathrm{SFT}}$, we further optimize the policy with GRPO:
\begin{equation}
    \theta^{\star} = \arg\max_{\theta} \mathcal{J}_{\mathrm{GRPO}}(\theta).
\end{equation}

This SFT$+$GRPO formulation aligns well with the structure of memory retrieval. SFT teaches the model to produce parseable reasoning traces, valid tool calls, and basic search behaviors, while GRPO sharpens relative preferences among competing retrieval trajectories under the same state. As a result, policy optimization can focus on higher-level decisions such as when to use parallel search, when to switch to serial search, how aggressively to denoise the evidence pool, and when the collected evidence is sufficient to stop.

\section{Model Architecture}

This section presents the architecture of MemRetriever, a ReAct-style memory retrieval model designed to construct a compact and sufficient evidence pool before answer generation. Unlike static top-$k$ retrieval, MemRetriever treats memory access as an explicit decision process: the model reasons about the current evidence state, chooses an appropriate retrieval or reflection action, observes the returned results, and iteratively updates the evidence pool until it is sufficient for the downstream reader.

\begin{figure}[t]
    \centering
    \smartincludegraphics{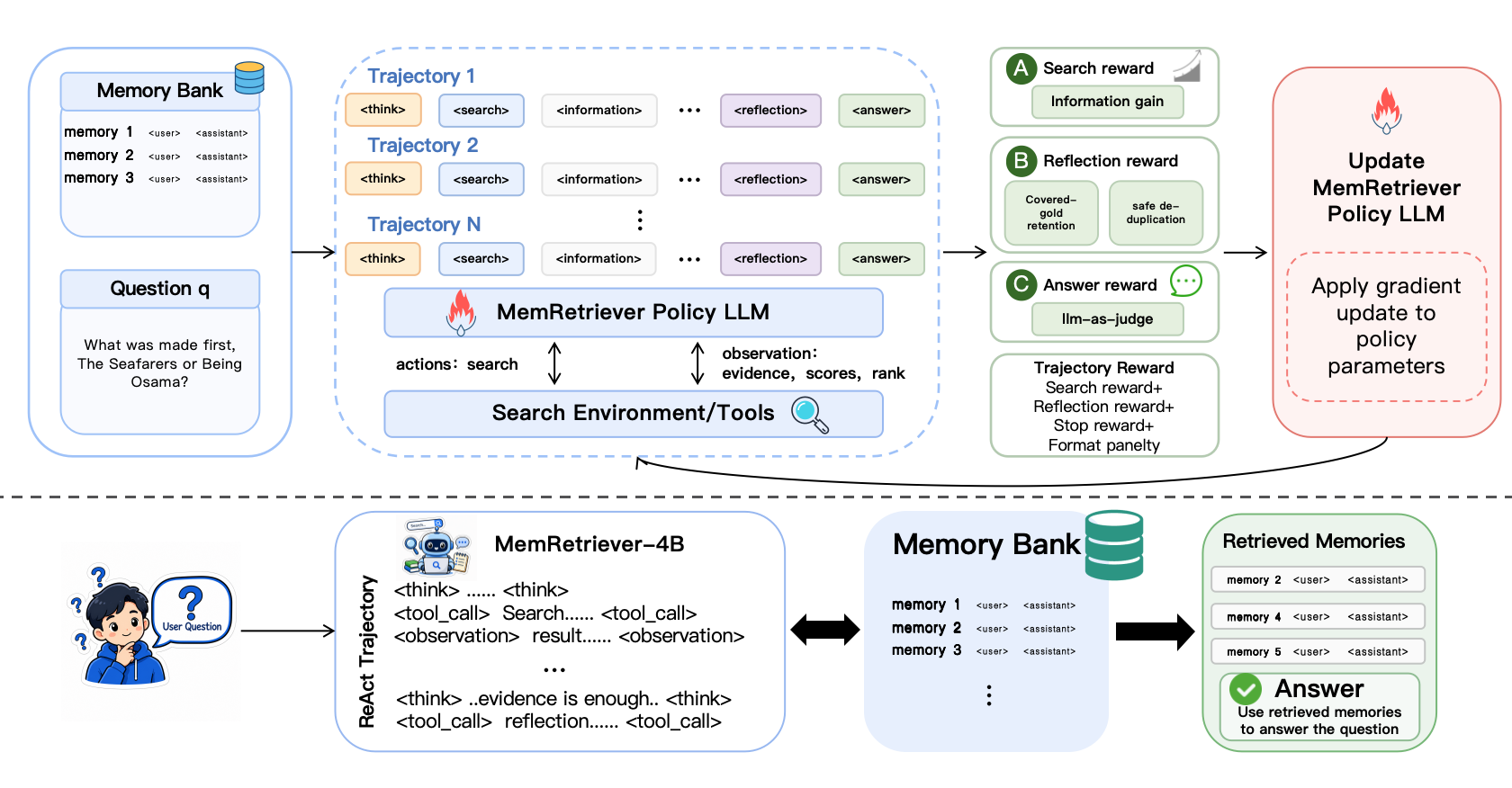}{0.95\linewidth}
    \caption{\textbf{Training and inference pipeline of MemRetriever.} The policy is first initialized through supervised warm start and then optimized with GRPO using search, reflection, stop, and format rewards. During inference, the trained search agent constructs an evidence pool through multi-step retrieval and denoising before passing it to the downstream reader model.}
    \label{fig:train_infre}
\end{figure}

\subsection{MemRetriever: A ReAct-Based Memory Retrieval Model}

MemRetriever is built on Qwen3-4B-thinking~\cite{qwen3} and trained under the ReAct (Reasoning + Acting) paradigm~\cite{react}. It has both internal thinking (\textit{Think}) and tool-calling (\textit{Action}) abilities. In our setting, the model is not responsible for directly producing the final user-facing answer. Instead, it serves as an intermediate search agent that actively retrieves, filters, and completes memory evidence for a downstream answering model.

The model design of MemRetriever focuses on four core questions:
\begin{itemize}
    \item \textbf{Q1 Search necessity:} Is the current evidence pool already sufficient to support downstream answering? Are key entities, relations, temporal clues, attributes, or comparison targets still missing and in need of further retrieval?
    \item \textbf{Q2 Search strategy selection:} Is the missing information open-ended and multi-directional, or has it already been localized to a specific information gap?
    \item \textbf{Q3 Evidence quality assessment:} Are the retrieved results relevant to the question? Can they support the reasoning chain? Do they contain noisy, redundant, or misleading evidence?
    \item \textbf{Q4 Stopping condition:} Is the current evidence sufficient to answer the question, or is the model trapped in repetitive search or a local retrieval state?
\end{itemize}

\paragraph{Tool-calling ability.}
Based on its internal reasoning, MemRetriever can invoke three retrieval-oriented tools. These tools correspond to the action space defined in Section~\ref{sec:problem} and are specialized for evidence acquisition and evidence purification:
\begin{itemize}
    \item \textbf{\texttt{parallel\_search}:} This tool is used when the evidence space is still unclear and the agent needs to expand recall. The model generates multiple complementary search queries in one step according to the current question and evidence state, exploring potentially relevant evidence paths from different perspectives.
    \item \textbf{\texttt{serial\_search}:} This tool is used when the missing information has been narrowed down to a specific information gap. The model issues a targeted query around one concrete missing fact, relation, entity, time point, or attribute.
    \item \textbf{\texttt{reflection\_and\_denoise}:} This tool is used to reflect on, filter, and denoise the current evidence pool. The model checks whether the retrieved evidence is truly relevant to the question, whether it can support a complete reasoning chain, and whether redundant, noisy, or off-topic evidence should be removed. It also judges whether the evidence pool is already sufficient for downstream answering; if so, the search loop terminates and only the key evidence is retained.
\end{itemize}

\paragraph{Overall inference process.}
The overall processing pipeline follows an \textit{input} $\rightarrow$ \textit{think} $\rightarrow$ \textit{action} (tool call) $\rightarrow$ \textit{observation} (tool result) loop. Given a user question and a long-term memory store, MemRetriever first analyzes the current evidence pool and identifies missing information. It then chooses between parallel search, serial search, and reflection-based denoising. The observation returned by the tool is added to or used to update the evidence pool, after which the model enters the next reasoning step. This loop continues until the reflection module determines that the evidence is sufficient, at which point MemRetriever outputs the final memory evidence pool for the downstream reader model.

Through this design, MemRetriever moves memory retrieval beyond one-shot similarity ranking. Parallel search improves recall under ambiguous evidence needs, serial search enables precise completion of identified gaps, and reflection-based denoising improves evidence quality by removing noise and redundancy. As a result, the downstream reader receives a more compact, relevant, and reasoning-ready evidence pool.

\section{Training}

This section describes the training pipeline of MemRetriever. The goal is to train a ReAct-style memory retrieval model that can decide whether to continue searching, what to search for, which tool to call, how to interpret memory observations, and when the collected evidence is sufficient for downstream answering. The overall training procedure consists of ReAct trajectory construction, supervised warm start, and GRPO-based reinforcement learning.

\subsection{ReAct Data Construction}

To train the ReAct capability of MemRetriever, we design a dedicated data construction pipeline for search-memory trajectories. For each sample, the model needs to reason over the current state, decide whether additional retrieval is necessary, identify the missing information, select an appropriate tool, and judge whether the current evidence pool can support the answer. We characterize this process with three actions: \texttt{parallel\_search}, which expands recall from multiple complementary directions when the evidence space is unclear; \texttt{serial\_search}, which performs targeted completion for a specific missing information point; and \texttt{reflection\_and\_denoise}, which reflects on the current evidence pool, removes noise, judges sufficiency, and scores and ranks the retained key evidence.

\paragraph{Construction method.}
We use diverse datasets as training sources and employ GPT-4.1-mini as a teacher model with strong reasoning ability. For each retrieval round, the teacher generates a complete Think--Action--Observation trajectory. Under the constraints of the system prompt, the teacher decides which tool to invoke according to the current state and outputs its internal reasoning through \texttt{<think>} tags. For search actions, we connect the teacher to a real vector retrieval backend based on Milvus, where bge-m3 is used as the embedding model. This allows the teacher to obtain realistic observation feedback during trajectory generation and ensures that the search $\rightarrow$ observation chain is logically valid.

During data construction, we mainly encountered three challenges. First, as the number of search rounds increases, the retrieval direction may drift away from the initially reasonable search logic. We observe that the reasoning path in the first-round \texttt{think} step is usually the most accurate, so we introduce a lightweight plan prompt as a guiding signal to constrain subsequent retrieval and reduce search drift. Second, when the model fails to retrieve the target information in one round, it may repeatedly search around the same or similar queries and become trapped in a local state. To mitigate this issue, we set a maximum number of search rounds and control invalid or low-gain repeated retrieval, thereby avoiding overly long trajectories and improving construction stability. Third, we filter out samples with overly simple search logic and further improve data diversity so that the trajectory distribution is more balanced. In the RL stage, we observe that overly simple rollouts in the SFT data can introduce bias and negatively affect subsequent policy optimization.

\paragraph{Data conversion and quality filtering.}
We unify the training data into ShareGPT format, alternating among \texttt{system}, \texttt{human}, \texttt{function\_call}, and \texttt{observation} messages. A trajectory is considered successfully constructed only when the model judges that the task can be answered and recommends \texttt{done} as the next action. We further apply format validation, JSON parsing checks, tool-call consistency checks, and retrieval-logic filtering to remove invalid or low-quality trajectories. The final dataset contains 3K SFT samples and 4K RL samples.

\subsection{Multi-Stage Training}

\paragraph{Stage 1 --- SFT warm-start.}
We first use the LLaMA-Factory framework~\cite{zheng2024llamafactory} to fully fine-tune Qwen3-4B-Thinking on the constructed search-memory ReAct trajectories. The main training configuration includes \texttt{learning\_rate=1.0e-5}, \texttt{lr\_scheduler\_type=cosine}, \texttt{num\_train\_epochs=3.0}, \texttt{enable\_thinking=true},\\\texttt{cutoff\_len=12288}.
Training uses DeepSpeed ZeRO-3 with \texttt{bf16=true}, \texttt{per\_device\_train\_batch\_size=1}, and \texttt{gradient\_accumulation\_steps=2}. This stage teaches the model the basic interaction protocol of the search-memory task, including generating valid reasoning traces, producing well-formed \texttt{<tool\_call>} outputs, understanding retrieved memory observations, and gradually collecting sufficient evidence through multi-round search. The SFT stage is necessary because reinforcement learning becomes substantially more stable once the model can already follow the required think--search--observe format.

\paragraph{Stage 2 --- GRPO reinforcement learning.}
We then adopt Group Relative Policy Optimization (GRPO) in the verl framework~\cite{sheng2024hybridflow} and conduct multi-round retrieval training. The main GRPO training configuration includes 
\texttt{train\_batch\_size=16}, 
\texttt{rollout.n=4}, 
\texttt{lr=1e-6}, 
\texttt{temperature=1.0}, 
\texttt{max\_assistant\_turns=6}, 
\texttt{max\_user\_turns=6}, 
and \texttt{max\_response\_length=6144}.We use GRPO because the reward signal in search tasks is trajectory-level and relative in nature. The model must decide the retrieval query, when to call tools, which evidence to retain or discard during reflection, and when to stop searching. The quality of these decisions cannot be reliably attributed to a single token; instead, it is better evaluated by the final effect of the complete search trajectory. In addition, the same question may admit multiple reasonable search paths, and the absolute reward scale can be affected by question difficulty, retrieval-store variation, and evidence distribution. The group-relative mechanism of GRPO normalizes comparisons within each question, reducing interference caused by reward-scale mismatch across samples and better matching the optimization objective of multi-step retrieval.

The design of our reward function further motivates the use of GRPO. The goal is to optimize whether the model can form a high-quality evidence acquisition strategy through multi-round interaction. During search, the model should discover new information relevant to the question; during reflection, it should retain evidence that truly supports the answer while removing distracting content; and at the end of the trajectory, it should stop when the evidence is sufficient and form an evidence set that can support correct answering. Therefore, the reward signal mainly comes from the overall trajectory effect, including whether retrieval brings new useful information, whether reflection compresses evidence while preserving key content, whether the final evidence is sufficient for answer generation, and whether tool invocation remains stable and executable. Such signals are sparse, discontinuous, and dependent on the combined effect of multiple decisions, making them difficult to decompose precisely into individual tokens or local actions. GRPO is therefore aligned with both the granularity and the optimization objective of our reward design, enabling the model to learn multi-step decision-making for evidence acquisition and evidence filtering.

\section{Experiments}

\subsection{Benchmarks}

We evaluate MemRetriever on five public benchmarks covering long-term memory retrieval and multi-hop question answering:
\begin{itemize}
    \item \textbf{LOCOMO}~\cite{locomo}: a long-dialogue memory benchmark with Single Hop, Multi Hop, Temporal, and Open Domain evaluation dimensions.
    \item \textbf{LongMemEval}~\cite{longmemeval}: a long-term memory-system benchmark covering single-session-preference, single-session-assistant, temporal-reasoning, multi-session, knowledge-update, and single-session-user tasks.
    \item \textbf{HotpotQA}~\cite{yang2018hotpotqa}: a classic multi-document multi-hop QA benchmark with sentence-level supporting-fact annotations. We randomly sample 200 examples from the dev set for evaluation.
    \item \textbf{MuSiQue}~\cite{trivedi2022musique}: a compositional multi-hop QA benchmark designed to better distinguish genuine multi-step reasoning from shallow pattern matching. We randomly sample 200 examples from the dev set for evaluation.
    \item \textbf{2WikiMultiHopQA}~\cite{ho2020constructing}: a multi-hop QA benchmark constructed from structured knowledge and Wikipedia text, with explicit evidence annotations. We randomly sample 200 examples from the dev set for evaluation.
\end{itemize}

Across all experiments, unless otherwise specified, MemRetriever uses the same retrieval configuration: \texttt{search-turns=6}, \texttt{serial-search-per-turn=6}, \texttt{final-evidence=5}, \texttt{parallel-search-queries-per-turn=3}, and \texttt{parallel-search-per-query=3}. These values are selected as empirically effective settings that provide strong retrieval and answering performance while keeping the search cost manageable. For the memory benchmarks, including LOCOMO and LongMemEval, the retrieval backend is the MemOS memory system~\cite{memos}, and MemRetriever calls the external \texttt{search memory} interface provided by MemOS to retrieve candidate memories. For LOCOMO and LongMemEval, both the answer model and the LLM-as-judge model are GPT-4.1-mini. For the QA benchmarks, the answer model is GPT-4o-mini and the judge model is GPT-4.1-mini. Higher values are better for all accuracy, retrieval, and judge metrics, while lower token consumption indicates better efficiency.

\subsection{LOCOMO Results}

Table~\ref{tab:locomo_retrieval} reports the retrieval and answer-quality results on LOCOMO. All agentic-search variants use the same three-tool pipeline with parallel search, serial search, and reflection-based denoising; the only difference is the policy model used to decide which tool to call and how to update the evidence pool. This allows us to compare MemRetriever with stronger general-purpose models under the same retrieval procedure. As shown in the table, reinforcement learning consistently improves the supervised warm-start model on both retrieval quality and downstream answer quality, while also reducing average token consumption. In particular, MemRetriever-4B-RL raises LLM-as-Judge accuracy from 0.785 to 0.830 compared with MemRetriever-4B-SFT, showing that the learned search policy improves the usefulness of the final evidence pool, not only its recall.

The gains mainly come from stronger constraints on the search process. The coverage-driven search reward encourages the model to discover new useful evidence, while the reflection reward based on final evidence coverage reduces noisy-memory accumulation and discourages the model from removing key evidence during compression. The lower token consumption is also a direct consequence of the three-tool agentic search design. In the final reflection action, the model explicitly removes evidence that it judges to be redundant or low-value before passing the final evidence pool to the downstream reader. As a result, the final pool usually does not reach the configured top-5 limit: stronger teacher models often retain only 2--3 key evidence items, while MemRetriever typically keeps about 3--4. This compression reduces the context length used for answer generation without discarding the evidence needed for correctness.

\begin{table}[H]
\centering
\caption{LOCOMO retrieval and answer-quality results. \texttt{token} denotes average token consumption.}
\small
\setlength{\tabcolsep}{4.2pt}
\begin{adjustbox}{max width=\linewidth}
\begin{tabular}{lcccccccc}
\toprule
\textbf{Method} & \textbf{hit@3\_all} & \textbf{hit@5\_all} & \textbf{ndcg@3} & \textbf{ndcg@5} & \textbf{recall@3} & \textbf{recall@5} & \textbf{token$\downarrow$} & \textbf{llm-as-judge} \\
\midrule
MemOS (embedding, Top-5) & 0.598 & 0.663 & 0.660 & 0.681 & 0.661 & 0.730 & 761 & 0.805 \\
Qwen3.7-Max & 0.671 & 0.746 & 0.745 & 0.764 & 0.739 & 0.810 & 767 & 0.827 \\
DeepSeek-v4-Pro & 0.684 & 0.758 & 0.758 & 0.778 & 0.755 & 0.825 & 650 & 0.847 \\
DeepSeek-v4-Flash & 0.699 & 0.762 & 0.752 & 0.773 & 0.759 & 0.825 & 664 & 0.837 \\
Gemini-3-Flash & 0.682 & 0.753 & 0.733 & 0.752 & 0.744 & 0.812 & 805 & 0.840 \\
MemRetriever-4B-SFT & 0.611 & 0.720 & 0.671 & 0.706 & 0.674 & 0.788 & 754 & 0.785 \\
MemRetriever-4B-RL & 0.667 & 0.750 & 0.685 & 0.719 & 0.726 & 0.806 & 673 & 0.830 \\
\bottomrule
\end{tabular}
\end{adjustbox}
\label{tab:locomo_retrieval}
\end{table}
\FloatBarrier

\subsection{LongMemEval Results}

Table~\ref{tab:longmemeval_retrieval} summarizes the LongMemEval results. As in the LOCOMO setting, all agentic-search methods follow the same three-tool retrieval pipeline and differ only in the model used as the search policy. The same pattern observed on LOCOMO also holds here: MemRetriever-4B-RL improves over the SFT model across the main retrieval metrics and achieves higher answer quality, while using fewer tokens on average. This suggests that reinforcement learning helps the model learn when additional retrieval is useful and when the current evidence pool is already sufficient.

MemRetriever-4B-RL also remains competitive with stronger closed-source models on several retrieval metrics, despite being a smaller specialized retrieval agent. These results further support the effectiveness of optimizing multi-step evidence acquisition directly, rather than relying only on supervised imitation of search trajectories.

\begin{table}[H]
\centering
\caption{LongMemEval retrieval and answer-quality results. \texttt{token} denotes average token consumption.}
\small
\setlength{\tabcolsep}{4.2pt}
\begin{adjustbox}{max width=\linewidth}
\begin{tabular}{lcccccccc}
\toprule
\textbf{Method} & \textbf{hit@3\_all} & \textbf{hit@5\_all} & \textbf{ndcg@3} & \textbf{ndcg@5} & \textbf{recall@3} & \textbf{recall@5} & \textbf{token$\downarrow$} & \textbf{llm-as-judge} \\
\midrule
MemOS (embedding, Top-5) & 0.754 & 0.840 & 0.961 & 0.968 & 0.861 & 0.915 & 472 & 0.772 \\
Qwen3.7-Max & 0.808 & 0.846 & 0.989 & 0.987 & 0.906 & 0.929 & 431 & 0.828 \\
DeepSeek-v4-Pro & 0.796 & 0.814 & 0.983 & 0.984 & 0.897 & 0.910 & 355 & 0.834 \\
DeepSeek-v4-Flash & 0.708 & 0.746 & 0.962 & 0.963 & 0.845 & 0.870 & 364 & 0.818 \\
Gemini-3-Flash & 0.796 & 0.834 & 0.975 & 0.975 & 0.895 & 0.920 & 432 & 0.836 \\
MemRetriever-4B-SFT & 0.728 & 0.792 & 0.968 & 0.961 & 0.856 & 0.893 & 702 & 0.766 \\
MemRetriever-4B-RL & 0.764 & 0.826 & 0.967 & 0.958 & 0.877 & 0.905 & 662 & 0.819 \\
\bottomrule
\end{tabular}
\end{adjustbox}
\label{tab:longmemeval_retrieval}
\end{table}
\FloatBarrier

\subsection{QA Results}

Table~\ref{tab:qa_results} reports results on HotpotQA, MuSiQue, and 2WikiMultiHopQA. Because prior systems report different evaluation protocols, we separate Exact Match (EM), F1, and LLM-as-Judge accuracy instead of forcing all methods into a single metric. A dash indicates that the corresponding metric is not reported or not directly comparable in the original setting.

\begin{table}[H]
\centering
\caption{QA results on three multi-hop benchmarks. EM and F1 are reported as ratios, while Judge is reported as a percentage. Dashes indicate unavailable or non-comparable metrics.}
\small
\setlength{\tabcolsep}{4.5pt}
\begin{adjustbox}{max width=\linewidth}
\begin{tabular}{llccc}
\toprule
\textbf{Dataset} & \textbf{Method} & \textbf{EM} & \textbf{F1} & \textbf{llm-as-judge} \\
\midrule
\multirow{6}{*}{HotpotQA}
& Search-R1/Qwen2.5-7B-Base & 0.430 & -- & -- \\
& ReSearch/Qwen-32B-Instruct & -- & -- & 63.6 \\
& MemSearcher/Qwen2.5-7B & 0.510 & -- & -- \\
& R1-Searcher/Qwen2.5-7B-Instruct & 0.440 & -- & -- \\
& MemRetriever-4B-SFT & 0.500 & 0.640 & 78.0 \\
& MemRetriever-4B-RL & \textbf{0.540} & \textbf{0.690} & \textbf{85.0} \\
\midrule
\multirow{6}{*}{MuSiQue}
& Search-R1/Qwen2.5-7B-Base & 0.196 & -- & -- \\
& ReSearch/Qwen-32B-Instruct & -- & -- & 33.4 \\
& MemSearcher/Qwen2.5-7B & 0.258 & -- & -- \\
& R1-Searcher/Qwen2.5-7B-Instruct & 0.160 & -- & -- \\
& MemRetriever-4B-SFT & 0.340 & 0.430 & 49.5 \\
& MemRetriever-4B-RL & \textbf{0.370} & \textbf{0.450} & \textbf{50.5} \\
\midrule
\multirow{6}{*}{2Wiki}
& Search-R1/Qwen2.5-7B-Base & 0.382 & -- & -- \\
& ReSearch/Qwen-32B-Instruct & -- & -- & 54.2 \\
& MemSearcher/Qwen2.5-7B & 0.490 & -- & -- \\
& R1-Searcher/Qwen2.5-7B-Instruct & 0.510 & -- & -- \\
& MemRetriever-4B-SFT & 0.520 & 0.580 & 64.0 \\
& MemRetriever-4B-RL & \textbf{0.550} & \textbf{0.610} & \textbf{67.0} \\
\bottomrule
\end{tabular}
\end{adjustbox}
\label{tab:qa_results}
\end{table}

Overall, MemRetriever performs strongly across all three multi-hop QA benchmarks, with especially clear gains on tasks requiring complex evidence composition. On MuSiQue, MemRetriever-4B-RL achieves the best EM, F1, and judge score among the listed methods. Since MuSiQue typically requires more reasoning steps and longer evidence chains, this result suggests that MemRetriever can progressively accumulate and preserve evidence that is useful for final answer generation, rather than merely retrieving locally similar passages.

We compare MemRetriever with four representative search-augmented baselines. Search-R1~\cite{jin2025searchr1} applies reinforcement learning to train LLMs to interleave reasoning and search-engine queries during rollout. ReSearch~\cite{chen2503research} learns when and how to invoke search as part of the reasoning chain through GRPO without supervised search trajectories. MemSearcher~\cite{yuan2025memsearcher} introduces memory management into a ReAct-style search agent to reduce accumulated history length and noisy context. R1-Searcher~\cite{song2025r1searcher} trains a general search-augmented reasoning policy with outcome-based reinforcement learning. Compared with these methods, MemRetriever focuses on the retrieval decision layer and explicitly separates parallel search, serial search, and reflection-based denoising. This design leads to stronger results on HotpotQA and MuSiQue, while remaining competitive on 2WikiMultiHopQA, suggesting that multi-step evidence control is especially useful when evidence composition becomes more complex.

These observations are consistent with our reward design. The search reward encourages the model to discover new valid information, the reflection reward reduces irrelevant evidence while preserving gold coverage, and the termination reward directly optimizes final answer sufficiency. Therefore, the learned policy improves not only retrieval recall, but also the usefulness of the acquired evidence for downstream multi-hop reasoning.

\subsection{Analysis}

Across LOCOMO, LongMemEval, HotpotQA, MuSiQue, and 2WikiMultiHopQA, the results show that MemRetriever brings stable improvements in both long-term memory retrieval and complex question answering. The benefits are particularly clear on multi-hop reasoning tasks: as reasoning chains become longer and the number of required evidence pieces increases, the combination of active search and reflection becomes more important.

The results also demonstrate that the proposed reward functions are effective. Coverage-driven search rewards help the model discover key evidence, recall-oriented reflection rewards reduce irrelevant information while preserving useful evidence, and judge-based termination rewards align the final evidence pool with answer sufficiency. Together, these rewards guide the model toward search behavior that better matches the downstream task objective.

MemRetriever remains beneficial across different retrieval backends and answer-model settings. Rather than depending on a specific reader model, it learns a general strategy for memory search and evidence management. This makes MemRetriever suitable as an independent intermediate module that can be integrated into existing retrieval-augmented generation systems to improve both evidence discovery and final answer quality.

\section{Conclusion}

This paper presents MemRetriever, a dedicated agentic memory retrieval model for long-term agent memory. Instead of relying on a static top-$k$ retrieval paradigm, MemRetriever formulates memory access as a multi-step search process in which the model actively plans retrieval, expands or narrows search directions, reflects on the accumulated evidence pool, filters noisy memories, and decides when the evidence is sufficient for downstream answering. This design turns memory retrieval from a one-shot matching problem into an explicit decision-making process over search actions and evidence management.

To train this capability, we construct ReAct-style search-memory trajectories and use them for supervised warm start, enabling the model to follow the required think--act--observe protocol and interact with retrieval tools in a stable format. We then optimize the model with GRPO, using reward signals aligned with the retrieval process: coverage-oriented search rewards encourage the discovery of new useful evidence, reflection rewards preserve key evidence while reducing noise, and termination rewards connect the final evidence pool to answer sufficiency. Together, these components allow MemRetriever to learn not only what to retrieve, but also how to retrieve, how to refine evidence, and when to stop.

Experiments on LOCOMO, LongMemEval, HotpotQA, MuSiQue, and 2WikiMultiHopQA demonstrate that MemRetriever improves both long-term memory retrieval and complex multi-hop question answering. The gains are especially clear in settings that require multiple pieces of evidence, longer reasoning chains, or dynamic evidence filtering. Compared with the SFT model, the RL-trained MemRetriever achieves better retrieval coverage and answer quality while reducing unnecessary token consumption on memory benchmarks, showing that the proposed reward design improves both effectiveness and efficiency.

More broadly, our results suggest that effective memory systems should not treat retrieval as a passive backend operation. Long-term agents need an intermediate decision layer that can actively control when to search, what to search for, how to filter retrieved memories, and when the current evidence is sufficient. Although this work focuses on long-term memory, the learned retrieval decision logic is storage-agnostic and can be applied to broader retrieval-augmented generation scenarios involving vector databases or external knowledge stores. Future work includes extending the tool set for richer evidence operations, improving robustness in online long-horizon interactions, and jointly optimizing retrieval, reflection, and answer generation in end-to-end agent systems.

\bibliographystyle{unsrt}
\bibliography{main}

\appendix

\section{Appendix A. Data Construction Prompt}

This appendix provides the system prompt used to construct ReAct-style search-memory trajectories for MemRetriever. The prompt defines the tri-tool interaction protocol, constrains the model not to answer the question directly, and specifies when to use \texttt{parallel\_search}, \texttt{serial\_search}, and \texttt{reflection\_and\_denoise} during trajectory generation.

\begin{promptbox}{Tri-Tool Search-Agent System Prompt}
\begin{lstlisting}[style=promptlisting]
SYSTEM_PROMPT_TRI_TOOL_AGENT = """You are an intelligent hybrid evidence-search agent for multi-hop question answering.

Your goal is to help a downstream answer model answer the question.
You must NOT answer the question yourself.

This is a single continuous conversation. All search actions, retrieved evidence, current-pool updates, scoring decisions, denoising decisions, and final conclusions happen in the same running message history.

You have exactly three tools:
1. parallel_search
2. serial_search
3. reflection_and_denoise

CRITICAL OUTPUT FORMAT REQUIREMENT:
Every response you generate MUST follow this exact two-part structure:

PART 1 (natural language, REQUIRED, never empty):
Write 2-4 sentences in plain text explaining:
- What you currently know and what is still missing.
- Why the chosen tool is the right one for this step.
- What you expect the tool to return and how it will advance the search.
Then, and only then, issue the tool call. The thought and the tool call together form one step. Do not skip the thought. Do not restate the tool arguments in the thought; explain the reasoning instead.

How to use the tools:
1. Use parallel_search when the evidence space is still unclear and you want to explore multiple complementary directions at once.
2. Use serial_search when you already know the main missing evidence slot and want one focused follow-up query.
3. Use reflection_and_denoise when you need to score the current_pool, remove noise, update which evidence remains in the current_pool, decide whether the evidence is sufficient, and choose the next action. You may call it in any round when scoring or pruning the current_pool would help.

Important rules:
1. Do not answer the question directly.
2. Do not use outside knowledge.
3. You do NOT need to call reflection_and_denoise after every search. Use it when it is helpful.
4. However, before stopping, you should usually use reflection_and_denoise so the final retained evidence is cleaned and summarized.
5. Use as few tool calls as necessary. Simple questions may need only 1-2 steps; complex ones may need more.
6. For parallel_search, provide 3-5 complementary queries.
7. For serial_search, provide exactly one focused query.
8. For reflection_and_denoise, assign usefulness scores to the current_pool items before deciding which ones to retain or discard.
9. Stop only when a downstream answer model could answer from the retained evidence without guessing.
10. If the current evidence is insufficient, reflection_and_denoise should say what is still missing and recommend either parallel_search, serial_search, or done.
"""
\end{lstlisting}
\end{promptbox}

\section{Appendix B. Case Study}

This appendix provides qualitative case-study rollouts that illustrate how MemRetriever performs multi-step evidence acquisition, denoising, and final evidence selection in both knowledge-intensive QA and personal memory retrieval settings.

\begin{casebox}{Case 1: Multi-hop knowledge retrieval \hfill \tikz[baseline=(tag.base)]{\node[fill=serialbg, text=serialtext, rounded corners=2pt, inner sep=2pt, font=\scriptsize\bfseries] (tag) {QA};}}

\begin{questionbox}
\textbf{Question:} The Russell Subdivision is a railroad line owned by a railroad headquartered in what state?
\end{questionbox}

\vspace{6pt}
\noindent\textbf{Step 1} \hfill \tagserial

\begin{thinkbox}
I first need to identify which railroad owns the Russell Subdivision. Once the owner is known, I can search for the headquarters location.
\end{thinkbox}

\noindent{\small\textbf{Query:} \texttt{Which railroad owns the Russell Subdivision?}}

\vspace{4pt}
\noindent{\scriptsize\textbf{Retrieved evidence:}}
\evretained{0}{The Russell Subdivision is a railroad line owned by CSX Transportation in the U.S. State of Kentucky.}
\evdiscarded{1}{The line runs from Russell, Kentucky to Greenup, Kentucky.}
\evdiscarded{2}{The Hudson Subdivision is a railroad line owned by CSX Transportation and leased by Amtrak in New York.}

\vspace{8pt}
\noindent\textbf{Step 2} \hfill \tagserial

\begin{thinkbox}
The first retrieval establishes CSX Transportation as the owner. The remaining missing information is the state where CSX is headquartered.
\end{thinkbox}

\noindent{\small\textbf{Query:} \texttt{Where is CSX Transportation headquartered?}}

\vspace{4pt}
\noindent{\scriptsize\textbf{Retrieved evidence:}}
\evretained{3}{CSX Transportation is headquartered in Jacksonville, Florida.}
\evdiscarded{4}{CSX Transportation is a Class I railroad in the United States.}
\evdiscarded{5}{CSX operates one of the three Class I railroads serving most of the East Coast.}

\vspace{8pt}
\noindent\textbf{Step 3} \hfill \tagreflect

\begin{thinkbox}
Evidence now supports the full chain: Russell Subdivision $\to$ CSX Transportation $\to$ Jacksonville, Florida. Retaining only the two key passages.
\end{thinkbox}

\noindent{\scriptsize\textbf{Final evidence pool:}}
\evretained{0}{Russell Subdivision is owned by CSX Transportation in Kentucky.}
\evretained{3}{CSX Transportation is headquartered in Jacksonville, Florida.}

\vspace{4pt}
\begin{answerbox}
{\scriptsize\textbf{Final answer (by working agent):}} \textbf{Florida.}
\end{answerbox}

\end{casebox}

\vspace{8pt}

\begin{casebox}{Case 2: Personal memory retrieval \hfill \tikz[baseline=(tag.base)]{\node[fill=parallelbg, text=paralleltext, rounded corners=2pt, inner sep=2pt, font=\scriptsize\bfseries] (tag) {Memory};}}

\begin{questionbox}
\textbf{Question:} What did Melanie realize after the charity race?
\end{questionbox}

\vspace{6pt}
\noindent\textbf{Step 1} \hfill \tagparallel

\begin{thinkbox}
The relevant memory may be phrased in different ways. I should issue several complementary searches around the race, what happened afterward, and Melanie's realization.
\end{thinkbox}

\noindent{\small\textbf{Queries:}}
\queryitem{q1: Melanie charity race realized after the race}
\queryitem{q2: charity race Melanie afterwards thought realized}
\queryitem{q3: Melanie race charity event conversation turned after race}

\vspace{4pt}
\noindent{\scriptsize\textbf{Retrieved evidence:}}
\evretained{0}{Melanie ran a charity race for mental health. She found it rewarding and reflected on the importance of mental health care.}
\evretained{1}{Melanie participated in a charity race for mental health. It prompted her to reflect on mental well-being and self-care.}
\evdiscarded{2}{Melanie participated in an advocacy event described as a cool experience with love and support.}
\evretained{3}{Caroline supported Melanie's participation and said raising awareness for mental health is rewarding.}

\vspace{8pt}
\noindent\textbf{Step 2} \hfill \tagreflect

\begin{thinkbox}
The pool contains direct evidence but also one noisy memory about a different advocacy event. Retaining memories about the charity race; discarding the unrelated event.
\end{thinkbox}

\noindent{\scriptsize\textbf{Final evidence pool:}}
\evretained{1}{Melanie participated in a charity race for mental health and reflected on mental well-being and self-care.}
\evretained{0}{Melanie found the charity race rewarding and reflected on the importance of mental health care.}
\evretained{3}{Caroline supported Melanie and emphasized that raising awareness for mental health is rewarding.}

\vspace{4pt}
\begin{answerbox}
{\scriptsize\textbf{Final answer (by working agent):}} \textbf{Melanie realized the importance of mental health care, mental well-being, and self-care.}
\end{answerbox}

\end{casebox}

\begin{center}
\captionof{figure}{Case study rollouts. \textbf{Case 1} demonstrates multi-hop knowledge retrieval via two serial searches followed by a reflect step. \textbf{Case 2} demonstrates personal memory retrieval via parallel search followed by a reflect step. Green entries denote retained evidence; struck-through gray entries denote discarded evidence.}
\label{fig:case-study}
\end{center}

\end{document}